\documentclass[12pt]{iopart}
\usepackage{graphicx}
\usepackage{epsfig}
\usepackage{epstopdf}

\begin{document}
\bibliographystyle{unsrt}
\title{Nonequilibrium quantum dynamics of atomic dark solitons}

\author{A D Martin and J Ruostekoski}
\address{School of Mathematics, University of Southampton, Southampton SO17 1BJ, United Kingdom}

\eads{\mailto{a.d.martin@soton.ac.uk},\mailto{janne@soton.ac.uk}}

\begin{abstract}
We study quantum dynamics of bosonic atoms that are excited to form
a phase kink, or several kinks, by an imprinting potential in a one-dimensional trap. We calculate dissipation due to quantum and thermal fluctuations in soliton trajectories, collisions and
the core structure. Single-shot runs show weak filling of a soliton core, typically deeper solitons in the case of stronger fluctuations and spreading/disappearing solitons due to collisions. We also analyze a soliton system in an optical lattice that shows especially strong fluctuation-induced phenomena.
\end{abstract}


\section{Introduction\label{Sec:Intro}}

Improved technology to manipulate ultra-cold atomic gases by optical lattice potentials \cite{Morsch},
atom chips \cite{Folman_AMOP_2002}, magnetic fields, etc., has provided tools to trap atoms in strongly confined
geometries and to adjust the dimensionality of the system. Such many-particle systems only interact
weakly with their environment and, over short time-scales, they may in many situations be approximately
considered as isolated and studied by models based on unitary quantum evolution. Moreover, the many-body atomic states may be engineered and controlled
at high accuracy and the detailed dynamics of atoms in response to external perturbations can be
experimentally investigated.

In a tightly-confined one-dimensional (1D) tube-potential quantum fluctuations of bosonic atoms can become strong \cite{Tolra_PRL_2004,Kinoshita_Nature_2006,Hofferberth_NPhys_2008}. The quantum effects may be further
enhanced, e.g., by applying an optical lattice potential along the axial direction of the trap \cite{Paredes_Nature_2004,Fertig_PRL_2005,Mun_PRL_2007}.
Stochastic phase-space methods provide a useful description of dissipative bosonic atom dynamics
in 1D traps due to quantum and thermal fluctuations of the atoms when the particle number is not
too low \cite{Isella_PRA_2005,Isella_PRL_2005,Isella_PRA_2006,POL03c}. The approximate quantum dynamics of atoms can be modelled in the truncated Wigner
approximation (TWA) \cite{Drummond_EPhysLett_1993,Steel_PRA_1998,Sinatra_JPhysB_2002,Blakie_AdvInPhys_2008} by unitary quantum evolution where the quantum statistical correlations
of the initial state may be accurately synthesized for a variety of quantum states
in the Wigner representation. The method can incorporate a very large phase-space, with a large number of degrees of freedom, in which case the atom dynamics, e.g., does not need to be restricted to the lowest energy band of an optical lattice. Dissipative dynamics emerges from a microscopic treatment of the unitary quantum evolution without any explicit damping terms and there is no need for the frequently problematic separation of quantum dynamics to `system' degrees of freedom and `environment' \cite{Anglin_NaturePhys_2008}. In addition, it is possible to study excitations of the system far from thermal equilibrium, quantum expectation values of various experimental observables and outcomes of single-shot measurements. Previous unitary TWA simulations \cite{Isella_PRL_2005} were qualitatively able to produce the experimentally observed damping rate of a dipolar centre-of-mass motion of bosonic atoms in
a very shallow, strongly confined 1D optical lattice \cite{Fertig_PRL_2005}. It was found that, due to quantum fluctuation induced momentum uncertainty, a small atom population reaches a critical velocity, representing
an onset of a dynamical instability in the corresponding classical system. The calculated damping
rate was approximately proportional to the atom population in the dynamically unstable velocity region.

Here we study dissipative quantum dynamics and relaxation of a 1D bosonic atomic gas that has been
excited to form a moving phase kink by an optical imprinting potential. Phase kinks have been experimentally imprinted in this way in 3D atomic Bose-Einstein condensates (BECs) by imaging
the atom cloud, e.g., through an absorption plate \cite{Burger_PRL_1999,Denschlag_Science_2000} or by
using optical holograms \cite{Becker_Nature_2008}. We recently showed how such a set-up could be
used to study dynamics of dark solitons in situations where quantum fluctuations are
important \cite{Martin_unpublished_2009}. The integrability of the soliton dynamics is broken by a trap and quantum and thermal fluctuations promote sound wave emission that may dissipate and eventually equilibrate with the soliton. Numerically tracking
soliton coordinates in individual stochastic realizations provided us with a tool to calculate quantum mechanical expectation values and uncertainties of the soliton position. We found, e.g., a surprising result that
the quantum expectation value of the speed of a soliton is {\em reduced} due to enhanced quantum fluctuations,
as a result of the nonlinear dependence of the soliton speed upon its phase distribution.
Single-shot runs in an optical lattice revealed effects of dynamical instabilities,
such as jittering oscillatory motion, splitting and disappearing solitons. In this paper, we study
a detailed structure of a quantum soliton, its quantum statistics and dissipative dynamics due to quantum and thermal fluctuations in fairly large atom-number systems. In particular, we consider
the effects of dissipative dynamics on a
soliton core structure, predicting a distinct difference between the filling behaviour of a soliton core in single-shot experimental runs and in the quantum expectation value of atom density found by averaging over many experimental runs. Single-shot realizations show only weak filling of a soliton core at later times and typically deeper solitons for the case of stronger quantum fluctuations. On the other hand, quantum expectation values for the atom density and pair-correlation functions are smeared out due fluctuation-induced drifting of solitons along the trap -- this is particularly visible for very slow solitons in an optical lattice system we consider. We also simulate quantum effects of collisions of soliton pairs and relaxation in a system of several solitons that may behave chaotically in the corresponding classical case. As in \cite{Martin_unpublished_2009} quantum and thermal fluctuations are synthesized within the truncated Wigner approximation in the quasi-condensate description. Dark solitons in nonlinear optical fibers were previously modelled using TWA \cite{optsoliton}. In atomic BECs, finite temperature dissipation has now been theoretically shown to manifest itself in an increasing oscillation amplitude of solitons in a harmonic trap \cite{Fedichev_PRA_1999,Jackson_PRA_2007,Cockburn_unpublished_2009,Martin_unpublished_2009}. Quantum properties of dark solitons \cite{Anglin_NaturePhys_2008} have recently started attracting considerable interest, including studies, e.g., of soliton core structure \cite{Dziarmaga_PRA_2002,Dziarmaga_PRA_2002_b,Dziarmaga_JPhysB_2003,Dziarmaga_PRA_2004,Damski_PRA_2006,Mishmash_PRL_2009,Mishmash_unpublished_2009},
phase kinks in a uniform space close to the Mott transition \cite{KRU09}, and
soliton statistics \cite{Negretti_PRA_2008_b}.

\section{Dark soliton experiments}

Dark solitons have been prepared in many experiments, and their subsequent dynamics probed. The first dark solitons in ultra-cold atomic Bose gases were generated by imprinting a sharp phase jump across the gas by optical potentials \cite{Burger_PRL_1999,Denschlag_Science_2000}, or by imprinting density defects using slow light pulses \cite{Dutton_Science_2001,Ginsberg_PRL_2005}. Subsequently, dark solitons have been created, e.g., by merging of two BECs \cite{Jo_PRL_2007} and by superfluid flow around a barrier \cite{Engels_PRL_2007}. In typical experiments the atoms were confined in 3D traps, or in 1D traps with only a weak radial trapping frequency, and the systems are accurately described by the classical Gross-Pitaevskii equation (GPE). In the phase imprinting method, a soliton is generated by applying a constant `light-sheet potential', of value $V_{\phi}$, to a part of the atom cloud, for time $\tau$ \cite{Burger_PRL_1999,Denschlag_Science_2000,Becker_Nature_2008}. The light-sheet potential is obtained by shining a far-detuned laser, e.g., through an absorption plate, so that the resulting dipole potential for the atoms exhibits a sharp edge. In the classical GPE limit, the potential imprints a phase jump
\begin{equation}
\phi_{c}=V_{\phi}\tau/\hbar,
\end{equation}
generating a dark (or grey) soliton (phase kink) with velocity ${\bf v}$ and the density dip at the phase kink (soliton core) with the minimum density $\rho_{s}$ \cite{Kivshar_PhysRep_1998}
\begin{eqnarray}
{\bf v}/c&=\cos(\phi_c/2),\\
\rho_{s}&=\rho_{b}\cos^{2}(\phi_{c}/2),\label{Eq_vn}
\end{eqnarray}
where $c$ is the speed of sound and $\rho_{b}$ is the density of the atomic background. The soliton
is stationary (dark) for $\phi_c=\pi$, with a zero density at the kink.
Other phase jumps produce moving (grey) solitons, with non-vanishing
densities at the soliton core, so that $|{\bf v}|\rightarrow c$ for $\phi_c\rightarrow 0$.
Soliton imprinting is accompanied by a density (sound) wave moving in the opposite direction to the soliton with speed approximately equal to $c$ \cite{Burger_PRL_1999}, created as a by-product of the perturbation of density by the imprinting potential.

More complex defect structures may generally be imprinted on superfluids using optical phase holograms
to shape laser fields, so that, via coupling with matter waves, the light acts as a hologram to
shape the BEC \cite{imprint}. A spatial light modulator was experimentally used to generate a multi-step potential and to prepare multiple phase kinks \cite{Stellmer_PRL_2008}. The potential
\begin{equation}
V_{i}(x,t)=\sum_{j}V_{\phi j}\theta(\tau_{j}-t)\theta(x-x_{j}),\label{Eq:VInt}
\end{equation}
(where $\theta$ is the Heaviside step-function) imprints phases $\phi_{j}$ at position $x_{j}$. The potential $V_{\phi_{j}}$ may be negative, such that $V_{\phi_{j}}=-V_{\tilde{\phi_{j}}}$ imprints a phase of $2\pi-\tilde{\phi_{j}}$, corresponding to a soliton travelling in the direction of the negative $x$-axis. Hence the initial positions and velocities of solitons may be accurately controlled. Other experiments have created multiple solitons by colliding two BECs \cite{Shomroni_NPhys_2009,Weller_PRL_2008}.

Subsequent to their formation, single solitons in a harmonic trap oscillate at angular frequency $\omega/\sqrt{2}$, where $\omega$ is the frequency of the trap \cite{Busch_PRL_2000}. Although many soliton experiments were performed in harmonic traps, these had a 3D character \cite{Anderson_PRL_2001}, such that solitons could decay into a hybrid of vortex lines and rings (`snake' instability) \cite{Muryshev_PRA_1999}; consequently, lifetimes of solitons were limited to less than one oscillation period. However, some experiments suppressed the snake instability by increasing the radial trapping potential \cite{Becker_Nature_2008,Weller_PRL_2008}, permitting observations of oscillations at the predicted frequency \cite{Becker_Nature_2008}. Recently, experiments probed the interactions between dark solitons \cite{Becker_Nature_2008,Stellmer_PRL_2008,Weller_PRL_2008} or between dark and dark-bright solitons \cite{Becker_Nature_2008}. Collisions between two dark solitons are accompanied by position-shifts \cite{Huang_PRA_2001,Zakharov_JETP_1973}, which change the soliton oscillation frequency \cite{Weller_PRL_2008}. During experiments, solitons repeatedly collided up to seven times without losing their integrity \cite{Weller_PRL_2008}.

Although some soliton experiments were sufficiently 1D to suppress snake instability, residual radial dynamics affected the oscillation frequencies by approximately 3\%  \cite{Weller_PRL_2008}. In more tightly confined 1D traps the radial density fluctuations may be completely suppressed.  For example, in an atom transport experiment, a 2D optical lattice divided a BEC into an array of decoupled 1D tubes \cite{Fertig_PRL_2005} with radial and axial trap frequencies in each tube $\omega_{r}=2\pi\times38$kHz and $\omega_{r}=2\pi\times60$Hz, respectively, and about 70 atoms in the central tube. A wide variety of 1D trapping schemes are also provided on atom chips using electromagnetic fields  \cite{Folman_AMOP_2002}. In a tightly confined 1D limit, bosonic atoms become more strongly interacting at low atom density \cite{Kheruntsyan_PRL_2003}, characterized by the effective interaction parameter $\gamma_{\rm int}=m g/\hbar^2 \rho$, where $\rho$ is the 1D atom density. For increasing values of $\gamma_{\rm int}$, enhanced phase fluctuations destroy the long-range phase coherence across the quasi-condensate, but density fluctuations remain suppressed--eventually the atoms reach the fermionized  Tonks-Girardeau regime \cite{Paredes_Nature_2004,Kinoshita_Nature_2006}.

\section{Truncated Wigner approximation}

In our analysis, quantum and thermal fluctuations of atoms are approximately included within TWA
\cite{Drummond_EPhysLett_1993,Steel_PRA_1998,Sinatra_JPhysB_2002,Isella_PRA_2006,Blakie_AdvInPhys_2008}.
Our TWA formalism and noise generation closely follows  \cite{Isella_PRA_2006}, except that here we
fix the total atom number and use a quasi-condensate description for quantum statistical atom correlations \cite{Martin_unpublished_2009}.
We replace the quantum field operators $(\hat{\psi},\hat{\psi}^{\dagger})$ by the classical fields $(\psi_{W},\psi^{*}_{W})$, governed by a 1D unitary evolution
\begin{equation}
i\hbar \frac{\partial}{\partial
t}\psi_{W}=\big(-\frac{\hbar^{2}}{2m}\frac{\partial^{2}}{\partial x^{2}}+V +gN_{\mbox{\scriptsize tot}}
|\psi_{W}|^{2}\big)\psi_{W},
\label{GPE}
\end{equation}
where the interaction strength $g=2\hbar \omega_{r}a$, the $s$-wave scattering length $a$, the total number of atoms $N_{\mbox{\scriptsize tot}}$, and $V=V_{\mbox{\scriptsize ext}}+V_{i}$ is the external 1D potential including the time-dependent imprinting potential. \Eref{GPE} formally coincides with the
classical GPE, but here quantum and thermal fluctuations are sampled in the initial state by an ensemble of stochastic fields $\psi_{W}(x,0)$ in the Wigner representation. Initially, before the phase imprinting, the atoms are assumed to be in thermal equilibrium in a harmonic trap, or in a combined harmonic trap and optical lattice. In a tight 1D trap we expect significant phase fluctuations that we found particularly important to phase kinks \cite{Martin_unpublished_2009} and in order to incorporate these we introduce a quasi-condensate description for the field operator \cite{Mora_PRA_2003}
\begin{equation}
\hat{\psi}(x,0) =\sqrt{\rho_{0}(x)+\delta{\hat{\rho}}(x)}\exp(i\hat{\varphi}(x)).\label{Quasi}
\end{equation}
The density $\delta{\hat{\rho}}(x)$ and phase $\hat{\varphi}(x)$ operators may be written in the Bogoliubov-type expansion \cite{Mora_PRA_2003}, requiring  $\delta\hat{\rho}/\rho_{0}$ and $|\delta l\Delta\hat{\varphi} |$ to be much less than one, where $\delta l$ is the spacing on the numerical grid on which we calculate the operators and $\Delta\hat{\varphi}$ is the gradient of the phase operator across one gridpoint. Thus (for $j>0$)
\begin{eqnarray}
\hat{\varphi}(x) &=\frac{-i}{2\sqrt{\rho_{0}(x)}}\sum_{j}\left(\varphi_{j}(x) \hat{\alpha}_{j}-\varphi_{j}^{*}(x)\hat{\alpha}_{j}^{\dagger}\right),\label{phaseop}\\
\delta\hat{\rho}(x)  &= \sqrt{\rho_{0}(x)}\sum_{j}\left(\delta\rho_{j}(x)\hat{\alpha}_{j}+ \delta\rho_{j}^{*}(x)\hat{\alpha}_{j}^{\dagger}\right)\,,\label{densityop}
\end{eqnarray}
%
where $\varphi_{j}(x)=u_{j}(x)+v_{j}(x)$ and $\delta\rho_{j}(x)=u_{j}(x)-v_{j}(x)$ are given in terms of the solutions to the Bogoliubov equations
\begin{eqnarray}
\left(\frac{-\hbar^{2}}{2m}\frac{\partial^{2}}{\partial x^{2}}+V-\mu+2\bar N_{0}g|\psi_{0}|^{2}\right)u_{j}-\bar N_{0}g\psi^{2}_{0}v_{j}=\epsilon_{j}u_{j},\\
\left(\frac{-\hbar^{2}}{2m}\frac{\partial^{2}}{\partial x^{2}}+V-\mu+2 \bar N_{0}g|\psi_{0}|^{2}\right)v_{j}-\bar N_{0}g\psi^{*2}_{0}u_{j}=-\epsilon_{j}v_{j},
\label{Eq_Bogoliubov}
\end{eqnarray} where $\mu$ is the chemical potential.
Here $\rho_{0}=\bar N_{0}|\psi_{0}(x)|^{2}$ and $\psi_{0}(x)$ is the ground state wavefunction with $\bar N_{0}$ particles.

We fix the total atom number to $N_{\mbox{\scriptsize tot}}$ and allow the ground-state and excited-state populations, $N_0$ and ${N}_{\mbox{\scriptsize nc}}$, to fluctuate. In the Bogoliubov approximation, the expectation value of the excited state population ($j>0$)
\begin{equation}
\bar{N}_{\mbox{\scriptsize nc}}=\int dx \sum_j \left[ \langle\alpha_{j}^{\dagger}\alpha_{j}\rangle\left(|u_{j}(x)|^{2}+|v_{j}(x)|^{2}\right)+|v_{j}(x)|^{2}\right],\label{Eq_BogEx}
\end{equation}
where $\langle \hat{\alpha}_{j}^{\dagger}\hat{\alpha}_{j}\rangle=\bar{n}_{j}=
\left[\exp(\epsilon_{j}/k_{B}T)-1\right]^{-1}$. In order to sample the stochastic initial state of TWA
with the correct quantum statistics, the operators $(\hat{\alpha}_{j}^{\dagger},\hat{\alpha}_{j})$ in (\ref{phaseop}) and (\ref{densityop}) are replaced by complex variables $(\alpha_{j}^{*},\alpha_{j})$, which may be generated from the Wigner distribution for harmonic oscillators with energy $\epsilon_{j}$ at temperature $T$ \cite{Isella_PRA_2006}, resulting in a stochastic Wigner representation $(\varphi_W(x),\delta\rho_W(x))$ of phase and density operators. The stochastic initial state for
the time evolution \eref{GPE} then reads
\begin{equation}
{\psi}_W(x,0) =\sqrt{\rho_{0,W}(x)+\delta{{\rho}}_W(x)}\exp(i{\varphi}_W(x)),\label{twainitial}
\end{equation}
where $\rho_{0,W}$ will be specified shortly.
Due to the Wigner representation, the stochastic variables correspond to symmetrically ordered expectation values of operators $(\hat{\alpha}_{j}^{\dagger},\hat{\alpha}_{j})$
and, e.g., the expectation values of the mode occupations differ from those calculated in the Wigner representation by half an atom per mode: $\langle \hat{\alpha}_{j}^{\dagger}\hat{\alpha}_{j}\rangle=\langle \alpha_{j}^{*}\alpha_{j}\rangle_{W}-1/2
$, where the notation $\langle ... \rangle_{W}$ indicates the expectation values are calculated in the Wigner representation. The excited state population for each run therefore reads
\begin{equation}\label{depletion}
N_{\mbox{\scriptsize nc}}=\int dx \sum_j \left[ \left(|\alpha_{j}|^{2}-1/2\right)\left(|u_{j}(x)|^{2}+|v_{j}(x)|^{2}\right)+|v_{j}(x)|^{2}\right],
\end{equation}
fluctuating around the correct expectation value (\ref{Eq_BogEx}). We set the ground-state population for each stochastic realization as $N_{0}=N_{\mbox{\scriptsize tot}}-N_{\mbox{\scriptsize nc}}$, so we can set for the initial state in each run $\rho_{0,W}(x)=(N_{0}+1/2)|\psi_0(x)|^2$. Then $N_{0}$ fluctuates around $\bar{N}_{0}=N_{\mbox{\scriptsize tot}}-\bar{N}_{\mbox{\scriptsize nc}}$ as required.

In simulations we vary the ground-state depletion $\bar{N}_{\mbox{\scriptsize nc}}/N_{\mbox{\scriptsize tot}}$. At temperature $T=0$ we keep the nonlinearity fixed at $N_{\mbox{\scriptsize tot}}g=100\hbar\omega l$, but adjust the ratio $g/N_{\mbox{\scriptsize tot}}$ where $l=(\hbar/m\omega)^{1/2}$. This is tantamount to varying the effective interaction strength $\gamma_{\rm in}$. We also study the effects of thermal depletion by varying $T$, by fixing both $N_{\mbox{\scriptsize tot}}$ and $g$ and setting $g/N_{\mbox{\scriptsize tot}}$ to be
sufficiently small so that quantum fluctuations are not dominant.

After evolution of the stochastic fields $\psi_{W}$, we extract physical quantities from the ensemble by calculating their quantum mechanical expectation values. Due to the symmetric ordering of the expectation values in the simulation data, we need to transform these to the normally ordered expectation values, corresponding to typical physical measurements \cite{Isella_PRA_2005,Isella_PRA_2006}. We find that the soliton position measurements considered in this paper are unchanged in the normal-ordering of phonon-mode occupation numbers, but that densities, pair-correlations and number fluctuations are quantitatively affected by the ordering. Hence the soliton position coordinates in individual experimental runs can be accurately extracted from the Wigner density $|\psi_{W}|^{2}$.


\section{Isolated soliton in a harmonic trap}\label{Sec:Single}

\subsection{Damping and dissipation in oscillatory dynamics}

We study the imprinting of single phase kinks and their subsequent dynamics in a harmonic trap in TWA. In our previous study \cite{Martin_unpublished_2009} we showed how quantum expectation values and uncertainties could be calculated for the dynamical variables of the soliton. We ran ensembles of hundreds of realizations (e.g.\ 400), and numerically tracked the soliton position coordinates $x(t)$ from $|\psi_{W}|^{2}$ in each run. This allowed us to calculate, e.g., the quantum mechanical position expectation value $\langle x \rangle$ and uncertainty $\delta x=\sqrt{\langle x^{2}\rangle-\langle x\rangle^2}$. Based on our findings, we made, e.g., the following observations that are relevant to our present study: (i) soliton trajectories are damped by both quantum and thermal fluctuations, increasing the amplitude of soliton oscillation; (ii) quantum mechanical soliton position uncertainties increased as a function of time for systems with large quantum depletion $\bar{N}_{\mbox{\scriptsize nc}}/N_{\mbox{\scriptsize tot}}$--this increase was due to the initial uncertainty in soliton velocity and deviations in the soliton trajectories as they interacted with sound waves; (iii) one of our most dramatic findings was that the quantum expectation values for the speed of a soliton were lower in systems with large quantum depletion which is attributable to a broad distribution of phase jump values across the kink caused by quantum fluctuations. We mapped the phase kink of each trajectory in the ensemble to a corresponding speed using the classical formulae (\ref{Eq_vn}) and found that due to the negative curvature of the soliton speed $|\cos(\phi/2)|$, a symmetric phase distribution always leads to a lower quantum expectation value for speed $\langle|\cos(\phi/2)|\rangle$ than the speed given by the phase expectation value $|\cos(\langle\phi/2\rangle)|$. Consequently, a greater phase uncertainty leads to a lower expectation value for the soliton speed.

\begin{figure}[tbp]
\centering
\includegraphics[width=12.0cm]{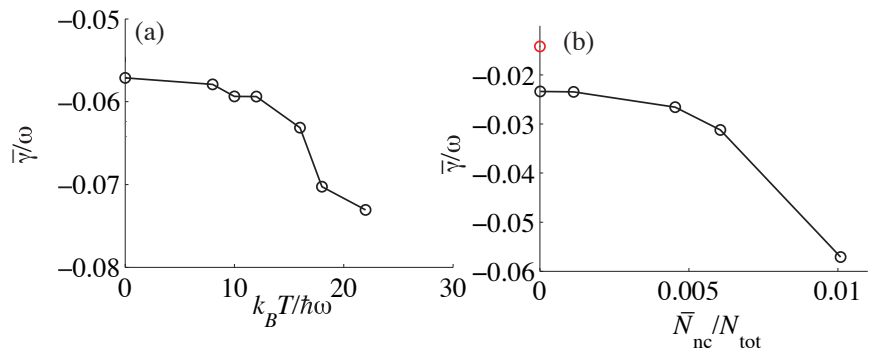}
\caption{(a) Quantum expectation value of the damping parameter $\bar{\gamma}$ of soliton oscillation amplitude in a harmonic trap  vs temperature for fixed $g N_{\mbox{\scriptsize tot}}=100\hbar\omega l$ and $N_{\mbox{\scriptsize tot}}=900$. $\bar{N}_{\mbox{\scriptsize nc}}/N_{\mbox{\scriptsize tot}}$ varies between $0.01$ at $T=0$, and $0.2$ at $T=22\hbar \omega/k_{B}$;  (b)  $\bar{\gamma}$ vs depleted fraction $\bar{N}_{\mbox{\scriptsize nc}}/N_{\mbox{\scriptsize tot}}$ at $T=0$ for fixed $g N_{\mbox{\scriptsize tot}}$ when we vary the effective interaction $g / N_{\mbox{\scriptsize tot}}$.  No points are plotted after $\bar{N}_{\mbox{\scriptsize nc}}/N_{\mbox{\scriptsize tot}}=0.01$ corresponding to $N_{\mbox{\scriptsize tot}}=900$ in which case fitting of the damped oscillations became less accurate. The Bogoliubov limit  $\bar{N}_{\mbox{\scriptsize nc}}/N_{\mbox{\scriptsize tot}}\rightarrow0$ corresponds to $\bar{\gamma}/\omega\simeq-0.023$ that is different from the classical GPE result $\bar{\gamma}/\omega\simeq-0.014$ (shown as an open red circle).
The damping increases ($\bar{\gamma}$ becomes more negative) faster than linearly with depletion. } \label{Damping_Plots}
\end{figure}

Here we evaluate in more detail the dissipation in soliton dynamics due to quantum and thermal fluctuations. We also study the structure of the soliton core and phase in individual stochastic realizations of TWA that represent possible outcomes of single-shot experiments. We then compare the quantum statistics of a soliton core in single-shot outcomes to quantum expectation values of the atom density which would be obtained in an experiment by averaging the density over many runs. We simulate the optical imprinting of phase kinks according to \eref{GPE} in the harmonic trapping potential $V_{\mbox{\scriptsize exp}}=m\omega^{2}x^{2}/2$, with the imprinting potential (\ref{Eq:VInt}), for $V_{\phi}=4166.7\hbar\omega$, applied over a period of $\tau=4.78\times10^{-4}/\omega$. Such a potential imprints a phase kink of $\phi_{c}=2$ in the classical (GPE) limit. Due to the stochastic sampling of quantum and thermal noise
in the initial state of each realization of the Wigner wavefunction, the actual phase jump values across the soliton core fluctuate from run to run, resulting in a variation in the shape of the solitons and in their initial velocities. After the imprinting the solitons oscillate in the harmonic trap.

Previously \cite{Martin_unpublished_2009}, we observed that soliton oscillations in a harmonic trap exhibit damping; since dark solitons have negative mass and kinetic energy, damping of a soliton increases its amplitude of oscillation. We track each soliton's trajectory by locating the local density minimum around the phase kink in $\psi_{W}$ for each realization. We fit individual trajectories with the curve $x(t)=f(t)\exp(-\gamma t)$, where $f(t)$ is a sinusoid and $\gamma<0$. We separately study the damping due to the ground-state depletion resulting from quantum fluctuations (at $T=0$) and from finite-temperature atoms (for parameters for which the corresponding $T=0$ damping is weak).
We only fit trajectories for parameters for which fluctuations are not too large, such that the trajectories are sufficiently close to sinusoidal. Figures \ref{Damping_Plots} (a) and (b) show that there is a faster-than-linear increase in damping with depletion in both cases.

\begin{figure}[tbp]
\centering
\includegraphics[width=12.0cm]{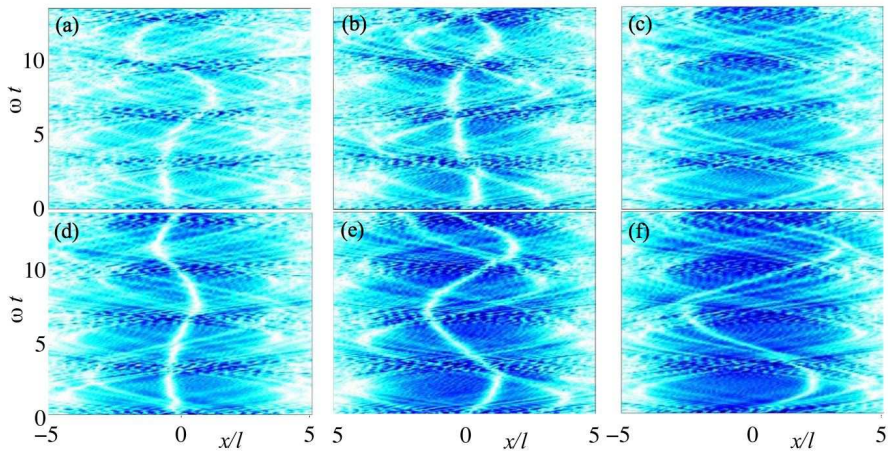}
\caption{Individual stochastic realizations of single soliton dynamics in a harmonic trap at $T=0$ for fixed nonlinearity $gN_{\mbox{\scriptsize tot}}=100\hbar\omega l$. In (a)-(c) the Wigner density $|\psi_{W}(x,t)|^{2}$ is plotted with $N_{\mbox{\scriptsize tot}}=50$ and the imprinted phase in the corresponding classical case $\phi_{c}=2.6$ (a) and $\phi_{c}=2.0$ [(b)-(c)]. (d)-(f) correspond to the same stochastic trajectories as (a)-(c) [the same realizations of $(\alpha_{j}^{*},\alpha_{j})$] but with $N_{\mbox{\scriptsize tot}}=100$. Soliton interactions with sound waves increase dissipation and oscillation amplitude. Dissipation is enhanced by strong quantum fluctuations [compare (a)-(b) to (d)-(e)].
} \label{Wigner_Plots}
\end{figure}

\begin{figure}[tbp]
\centering
\includegraphics[width=12.0cm]{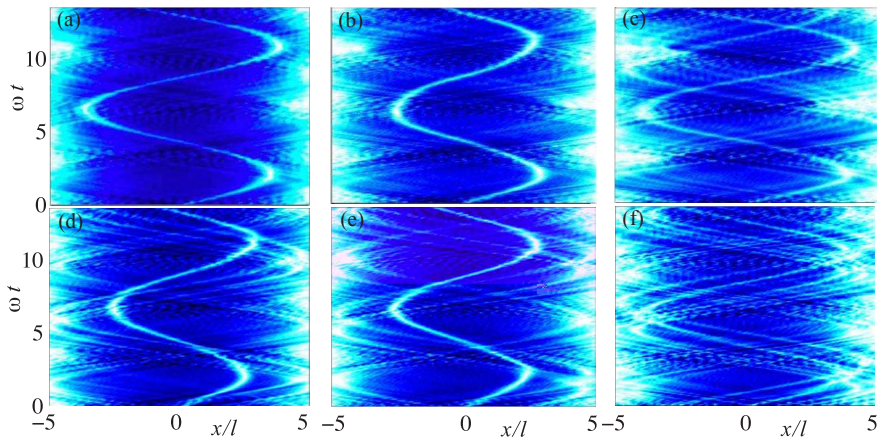}
\caption{Individual realizations of trajectories for a single imprinted soliton at different temperatures for $N_{\mbox{\scriptsize tot}}=900$, $gN_{\mbox{\scriptsize tot}}=100\hbar\omega l$ and $\phi_{c}=2.0$. The Wigner density $|\psi_{W}(x,t)|^{2}$ is plotted for (a) $T=0$, (b)-(c) $T=16\hbar\omega/k_{B}$, and (d)-(f) $T=22\hbar\omega/k_{B}$.
For large thermal depletion, sound wave emission is clearly visible, as in the case of large quantum depletion in figure \ref{Wigner_Plots}. Interactions between the sound waves and solitons are weaker than in figure \ref{Wigner_Plots}, even in cases with similar depletion: $\bar N_{\mbox{\scriptsize nc}}/N_{\mbox{\scriptsize tot}}\simeq 0.2$ in (d)-(f) as in figure \ref{Wigner_Plots}(a)-(c).} \label{Wigner_Plots_T}
\end{figure}

The source of this damping behaviour can be deduced by consideration of individual trajectories in ensembles with large depletion (shown in figures \ref{Wigner_Plots} and \ref{Wigner_Plots_T}). Strongly dissipative behaviour occurs when energy is exchanged between the solitons and sound waves. This is particularly clear for large quantum depletion shown in figures \ref{Wigner_Plots}(a)-(c). We find that at finite temperature large thermal depletion has an associated sound wave background [figure \ref{Wigner_Plots_T}(d)-(f)], but the soliton trajectories are less perturbed by the sound waves than in cases of large quantum depletion. We contend that the breaking of integrability by the harmonic trap causes the energy to disperse amongst the excitations. Solitons with relatively less negative kinetic energy are more prone to dissipation, and the damping rate is greatest in these trajectories.

The limit $g/N_{\mbox{\scriptsize tot}}\rightarrow0$, for fixed $gN_{\mbox{\scriptsize tot}}$, corresponds to $\bar{N}_{\mbox{\scriptsize nc}}/N_{\mbox{\scriptsize tot}}\rightarrow0$ and we expect the Bogoliubov approximation to become increasingly accurate (if the ground state and the excited state populations are not solved self-consistently, $\bar{N}_{\mbox{\scriptsize nc}}$ is approximately constant and $N_{\mbox{\scriptsize tot}}\rightarrow\infty$ \cite{Shrestha_PRA_2009}).
Surprisingly, in the limit that  $g/N_{\mbox{\scriptsize tot}}\rightarrow0$,  $\bar{\gamma}$ converges to a more negative value than that found in classical GPE simulations [figure \ref{Damping_Plots}(b)]. The classical soliton exhibits damping due to interaction with sound waves that originate from the imprinting process. Such interactions may represent dynamically unstable processes that require very weak numerical noise to be triggered which may be absent in GPE. Alternatively, the soliton trajectory may represent a state with non-classical statistics that does not converge to the classical GPE value, similarly to the atom number-squeezed states. Our simulations seem to indicate that the first case is a more likely explanation, since, e.g., an uncorrelated noise at each spatial grid point, with the magnitude $\sim 10^{-4}$ weaker than quantum
vacuum noise, reproduces the $g/N_{\mbox{\scriptsize tot}}\rightarrow0$ TWA limit. 


\subsection{Soliton core}

\begin{figure}[tbp]
\centering
\includegraphics[width=12.0cm]{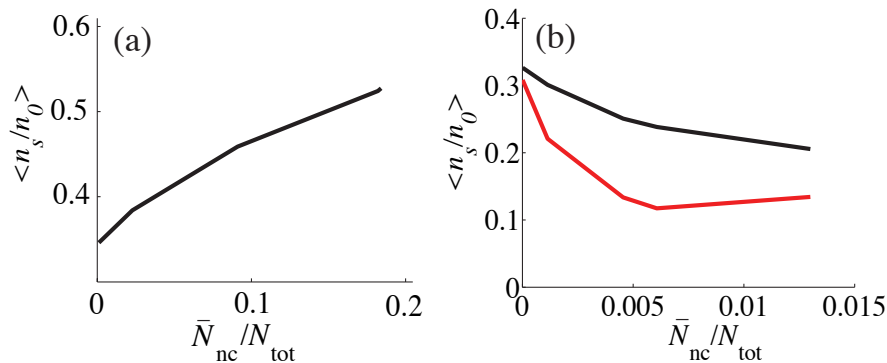}
\caption{
Quantum expectation value for atom density in soliton core divided by the background density $\langle \rho_{s}/\rho_{b}\rangle$ at $T=0$. (a) at time $t=6.7\times 10^{-3}/\omega$ (immediately after imprinting). Larger quantum fluctuations indicate a shallower soliton core;  (b) when the soliton crosses the point $x=-0.4 l$ for the first time (lower/red line) and for the second time (upper/black line). Larger quantum fluctuations now typically indicate a deeper soliton core that exhibits some filling as a function of time.
} \label{Filling_Plots}
\end{figure}

Many studies \cite{Law_PRA_2003,Dziarmaga_PRA_2002,Damski_PRA_2006,Dziarmaga_JPhysB_2003,Dziarmaga_PRA_2004} suggest a link between dissipative soliton dynamics and filling of a soliton core. To investigate this, we calculate the ratio $\rho_{s}/\rho_{b}$ of the minimum density in the soliton core to the background density around the soliton. For the classical soliton solution these variables are related by (\ref{Eq_vn}). We stress the distinction between the quantum expectation value of the soliton depth, which we infer from the expectation value $\langle \rho_{s}/\rho_{b}\rangle$ found by locating the soliton and evaluating its depth in each stochastic realization, and the value of the density notch in the quantum expectation value $\langle \rho(x)\rangle$, obtained by ensemble averaging over densities in many realizations. We first calculate the quantum expectation value $\langle \rho_{s}/\rho_{b}\rangle$ at time $t=6.7\times 10^{-3}/\omega$ -- very soon after the phase imprinting when the density dip is still forming and all solitons in different stochastic realizations are still approximately centered at $x=0$. We approximate $\rho_{b}$ by $\rho(0.39l)$ in each trajectory. Figure \ref{Filling_Plots}(a) shows how the filling of the emerging notch increases due to quantum fluctuations at $T=0$, as density fluctuations obscure the soliton. However, by time $t=6\times 10^{-2}/\omega$ after the formation of the density dip, the filling behaviour reverses and $\langle \rho_{s}/\rho_{b}\rangle$ becomes smaller in systems with large depletion.  This increase in soliton depth due to quantum fluctuations is consistent with the aforementioned slowing down of the soliton velocities found in our previous study \cite{Martin_unpublished_2009} in which case the increased phase uncertainty reduced the quantum expectation value of soliton speed $\langle|\cos(\phi/2)|\rangle$ due to the nonlinear dependence of the speed upon the phase jump. The classical formula between the phase jump and soliton core density \eref{Eq_vn} also exhibits a negative curvature and we expect strong quantum fluctuations to reduce the quantum expectation value of soliton core density, reflected by $\langle\cos^{2}(\phi/2)\rangle$. This deepening of solitons is surprising considering recent studies \cite{Law_PRA_2003,Dziarmaga_PRA_2002,Damski_PRA_2006} which suggest that a soliton core in individual runs may develop density peaks due to ground-state depletion, and we would naively expect shallower, faster solitons. Instead we find that the depth of a soliton in individual runs is dominated by large phase fluctuations which give rise to the opposite effect, so that, after the formation, a soliton core is deeper in systems with large depleted fraction.

We now consider the behaviour of the soliton depth during the evolution in a harmonic trap. Figure \ref{Filling_Plots}(b) shows the quantum expectation value $\langle \rho_{s}/\rho_{b}\rangle$ as the soliton passes the point $x=-0.4 l$ for the first and second time during its oscillation, which occur at times $t\simeq4.8$ and $t\simeq8$. In this case we may determine $\rho_{s}$ in every trajectory by the density at position $x$ at the time when the soliton passes, and $\rho_{b}$ by averaging the density at the point $x$ between consecutive passes of the soliton. We find that for individual trajectories, the soliton depth decreases for each pass; i.e., there is a slight filling of the soliton core with time. This is accompanied by a gradual speeding of the solitons due to dissipation ($\gamma<0$). Quantum fluctuations enhance filling of the soliton core. After formation, the soliton in systems with strong quantum fluctuations, however, is initially typically deeper than in the case of weak fluctuations, and figure \ref{Filling_Plots} shows that they remain deeper despite experiencing greater filling.

\begin{figure}[tbp]
\centering
\includegraphics[width=12.0cm]{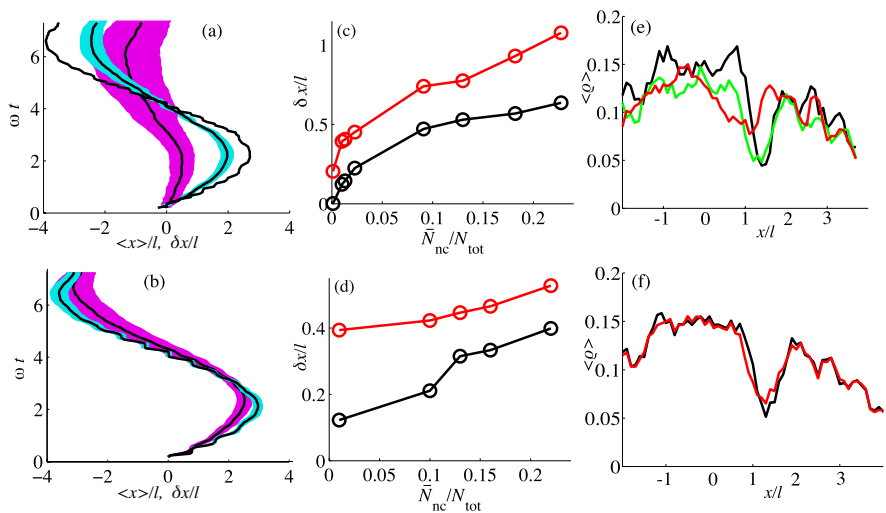}
\caption{(a) Quantum expectation value for the soliton position $\langle x(t)\rangle$ (lines) and its uncertainty $\delta x$ (shaded area) for solitons in a harmonic trap  at $T=0$ for (with increasing oscillation amplitudes) $N_{\mbox{\scriptsize tot}}=50$ (magenta), $N_{\mbox{\scriptsize tot}}=440$ (cyan) and $N_{\mbox{\scriptsize tot}}=8000$ (negligible width); (b) as in (a), but with $N_{\mbox{\scriptsize tot}}=900$, $T=22\hbar\omega/k_{B}$ (magenta) and $T=0$ (cyan);
(c) the soliton position uncertainty $\delta x$ at two fixed times $t_{1}\simeq1/\omega$ (lower/black line) and $t_{2}\simeq 3/\omega$ (upper/red line) at $T=0$, with $N_{\mbox{\scriptsize tot}}$ varying between 40 and 8000; (d) as in (c) but varying $T$ between 0 and $16 \hbar\omega/k_{B}$ for fixed $N_{\mbox{\scriptsize tot}}=900$;
(e)-(f) expectation value of atom density $\langle\rho\rangle$ at time $t\simeq3.7/\omega$, obtained from $|\psi_{W}|^{2}$ by converting from the Wigner to the normally-ordered representation; (e) at $T=0$ curves correspond to $N_{\mbox{\scriptsize tot}}=50$ (intermediate/red line), $N_{\mbox{\scriptsize tot}}=440$ (light/green line) and $N_{\mbox{\scriptsize tot}}=8000$ (dark/black line); (f) for fixed $N_{\mbox{\scriptsize tot}}=900$ with the curves corresponding to $T=0$ (dark/black line) and $T=16\hbar\omega/k_{B}$ (light/red line).
Increasing position uncertainty at low atom numbers create a filling effect when the density is averaged over many realizations. }\label{Width_Plots2}
\end{figure}

Numerically tracking the location of the phase kink in each individual realization, we have established that the expectation value for the depth of the soliton core is larger for larger values of $\bar{N}_{\mbox{\scriptsize nc}}/N_{\mbox{\scriptsize tot}}$. Solitons may drift around due to quantum fluctuations and in different runs they are generally not located in the same spatial position at the same time -- soliton position uncertainties grow as a function of time due to both initial velocity uncertainties and interactions of solitons with sound waves \cite{Martin_unpublished_2009}. In an ensemble averaged density over many single realizations, corresponding to the quantum
expectation value for the atom density $\langle\rho(x,t)\rangle$, soliton core density profiles appear broader and shallower due to different soliton locations in individual runs, as displayed in figure \ref{Width_Plots2}. The soliton position has a particularly large uncertainty in the presence of strong fluctuations and we find that, despite the fact that the expectation value of the soliton core is deeper in individual realizations with stronger quantum fluctuations, the density notches become shallower in $\langle\rho(x,t)\rangle$ when the fluctuations are enhanced. The flattening of  $\langle\rho(x,t)\rangle$ due to the wandering solitons is similar to the previous results obtained using a Bogoliubov analysis \cite{Dziarmaga_JPhysB_2003,Dziarmaga_PRA_2004} around a stationary dark soliton state showing that $\langle\rho(x,t)\rangle$ is smeared out due to fluctuations in the soliton positions.

Matrix product state simulation of a dark soliton in a bosonic atomic gas in a lattice in the tight-binding approximation with unit filling was studied in references \cite{Mishmash_PRL_2009,Mishmash_unpublished_2009}. From the expectation value of the atom density it was shown that an instantaneously imprinted phase kink with a vanishing density at the soliton core at the center of the lattice decays due to quantum fluctuations, as the soliton core gets filled with atoms. This is consistent with the aforementioned Bogoliubov analysis  \cite{Dziarmaga_JPhysB_2003,Dziarmaga_PRA_2004}. Moreover, based on the time-evolution of the atomic pair-correlation function it was argued that the soliton core also in single-shot experimental realizations is filled. In the simulations the initial state of $g_2(0,k)=\langle a^\dagger_k a^\dagger_0 a_0 a_k \rangle=0$, since the central site $j=0$ was empty (the location of the density dip of the soliton) and the other sites had approximately one atom. (Here $a_k$ denotes the annihilation operator for the atoms at the site $k$.) At later times $g_2$ became a flatter (and non-vanishing) function of $k$. Whether the filling of a soliton core in single-shot experiments in such a system duly happens, however, is inconclusive. This is because the evolution of $g_2$ can in many cases equally well be explained by a soliton whose core does not get filled in individual single-shot runs, but which randomly drifts along the lattice with the standard deviation of the soliton position increasing as a function of time. As a simple, idealized example, consider a lattice system where the long-range correlations may be ignored so that $g_2(0,k)\sim n_0 n_k$. Here $n_k$ represents the atom density at the site $k$ that is obtained by an ensemble average over many single-shot runs. The soliton is initially assumed to be located at the central site, with $n_0=0$ and $n_j=1$, for $j\neq0$, and $g_2(0,k)=0$ for all $k$. At some later time we may have $n_k=(L-1)/L$ and $g_2(0,k)\sim (L-1)^2/L^2$, for all $k$, where $L$ denotes the number of sites. This can represent
an outcome where every single-shot run yields a constant atom density $(L-1)/L$ for all $k$, but equally well a situation where each individual run has a single soliton at a random location along the lattice, so that the probability of finding a soliton (with a vanishing density) at an arbitrary site $k$ in each realization is $1/L$ and the probability of not finding a soliton at the same site $(L-1)/L$.

Our numerical simulations of quantum dynamics of a soliton with the corresponding classical value of the imprinted phase jump $\phi_c=\pi$ in a lattice demonstrate a similar phenomenon (as shown in Section \ref{Sec:Lattice}): Individual stochastic realizations exhibit drifting soliton dynamics along the lattice due to quantum fluctuations. The ensemble average of the atom density and the pair-correlation function $g_2$ become flatter as a function of time when the solitons have more time to drift over longer distances in the lattice. At the same time, however, the soliton cores in single-shot runs show very little, if any, effects of filling.

\section{Colliding solitons \label{Sec:Collisions}}

Recent studies of quantum dynamics of solitons in an optical lattice in the tight-binding limit with one mode function per lattice site have shown that soliton density may spread after colliding due to quantum fluctuations \cite{Mishmash_PRL_2009,Mishmash_unpublished_2009}. We examine soliton collisions in the absence of a lattice, and consider the relationship of such inelastic collisions with the soliton structure in the presence of strong quantum fluctuations.

\begin{figure}
\centering
\includegraphics[width=12.0cm]{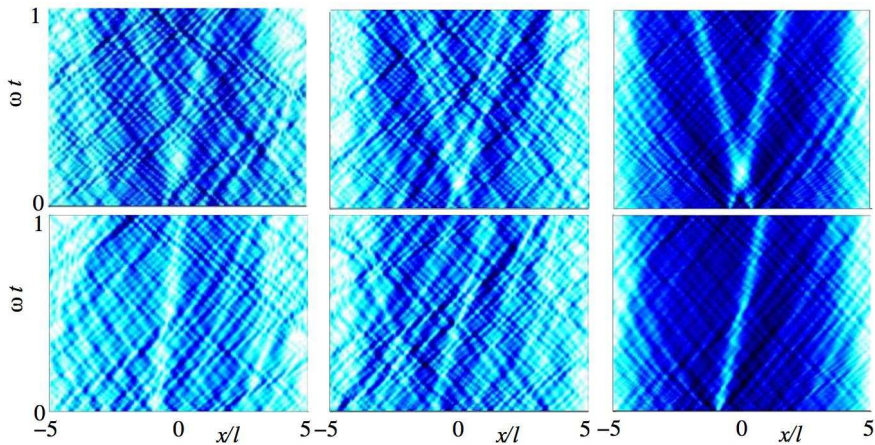}
\caption{Individual stochastic realizations of $|\psi_{W}(x,t)|^{2}$ for collisions of two solitons at $T=0$ (top row), and corresponding trajectories of single solitons (bottom row) for fixed $gN_{\mbox{\scriptsize tot}}=100\hbar\omega l$, with $N_{\mbox{\scriptsize tot}}=100$ (left and centre) and $N_{\mbox{\scriptsize tot}}=900$ (right). At low atom number, for the case of strong quantum fluctuations, the solitons split after the collision. Comparison between the top and bottom rows shows how quantum fluctuations turn collisions inelastic, resulting in splitting solitons.} \label{Two_Plots}
\end{figure}

We simulate the generation of two solitons using an optical imprinting potential (\ref{Eq:VInt}) with two steps $V_{\phi}=4166.7\hbar\omega$ at $x_{b}=-x_{a}=0.49 l$ which in a classical system would imprint phase kinks of $\phi_{c}=2$ and $\phi_{c}=2\pi-2$ with opposite velocities $\pm\cos(\phi_{c}/2)$. We study the effects of quantum fluctuations on the imprinting of the phase kinks and the subsequent collisions of two such kinks (at about $t\simeq 0.2/\omega$). As before, we do this by varying the effective interaction strength, corresponding to different values of the ground-state depletion, by keeping the nonlinearity $N_{\mbox{\scriptsize tot}}g$ fixed but
adjusting the ratio $g/N_{\mbox{\scriptsize tot}}$. Figure \ref{Two_Plots} shows $|\psi_{W}(x)|^{2}$ for individual realizations of soliton collisions, and, for comparison, trajectories with the same stochastic realization, but using an imprinting potential that imprints only one of the solitons. We observe splitting/spreading of the solitons as they emerge from collisions in the presence of large quantum fluctuations. Since the corresponding trajectories of single solitons do not exhibit this behaviour, we conclude that the splitting is induced by quantum fluctuations during the soliton collisions. Despite the spreading of the region of low atom density around the soliton core in individual realizations, the minimum density in the soliton core does not increase in these trajectories. We infer that one soliton splits into many solitons of comparable depth and velocity.

\begin{figure}
\centering
\includegraphics[width=12.0cm]{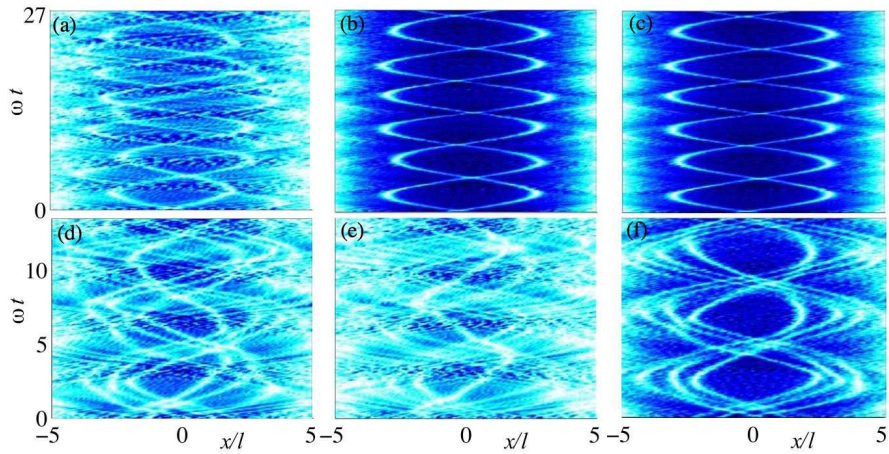}
\caption{Individual stochastic realizations of $|\psi_{W}(x,t)|^{2}$ for repeated collisions between different solitons. (a)-(c) Two solitons with (a)  $N_{\mbox{\scriptsize tot}}=100$ (b) $N_{\mbox{\scriptsize tot}}=1700$ and (c) $N_{\mbox{\scriptsize tot}}=8000$. (d)-(f) Six solitons with (d) and (e) $N_{\mbox{\scriptsize tot}}=100$, (f) $N_{\mbox{\scriptsize tot}}=900$. Trajectories (d)-(e) show dramatic perturbations of the soliton motion, causing some solitons in (e) to disappear.} \label{Two_Plots_Long}
\end{figure}

Due to the dipolar motion in a harmonic trap, two solitons can experience repeated collisions. Recent experiment \cite{Weller_PRL_2008} showed seven repeated collisions of two solitons generated by merging two BECs.  The parameters were close to those in our system: $\omega\simeq 333$Hz, $\omega_{r}\simeq 5592$Hz and $N_{\mbox{\scriptsize tot}}\simeq 1700$ ($gN_{\mbox{\scriptsize tot}}\simeq205\hbar\omega l$); or $\omega\simeq 364$Hz, $\omega_{r}\simeq 2563$Hz and $N_{\mbox{\scriptsize tot}}\simeq 950$  ($gN_{\mbox{\scriptsize tot}}\simeq50.1\hbar\omega l$). The experiment was performed in an elongated 1D trap, but without sufficiently strong transverse confinement that would have suppressed radial density fluctuations, and quantum fluctuations are not expected to be important. In our 1D system we find little quantum effects at similar atom numbers (other than a reduction in the soliton velocities).
However, after reducing the atom number to $N_{\mbox{\scriptsize tot}}=100$, we observe soliton trajectories becoming perturbed [figure \ref{Two_Plots_Long}(a)] or disappearing after several collisions that may be detectable in an experiment using a tighter transverse confinement than the one in  \cite{Weller_PRL_2008}.

Increasing the number of solitons from two introduces the possibility of far richer dynamics. Systems of three or more harmonically trapped {\em bright} solitons may exhibit chaotic dynamics; the dynamics are not integrable due to the interplay between the harmonic motion and the soliton interaction \cite{Martin_PRL_2007}. The similarity between the interactions of bright solitons and those of dark solitons suggests that multiple dark solitons in a harmonic potential may also display chaotic dynamics. Multiple soliton collisions may also appear, e.g., in self-focusing and revival dynamics of BECs \cite{Ruostekoski_PRA_2001}. As a signature of quantum fluctuations, we find that in such systems, solitons are more susceptible to disappearing. We use the imprinting potential (\ref{Eq:VInt}) to imprint six phase kinks all with different initial positions and velocities. Figure \ref{Two_Plots_Long}(d)-(f) shows   $|\psi_{W}(x)|^{2}$ for individual realizations. The enhanced dissipative effects visible in  \ref{Two_Plots_Long}(d)-(e) where solitons trajectories become rapidly perturbed or disappear, may be related to dynamical instability of the system associated with chaotic dynamics, or merely due to the increased number of collisions experienced by each soliton in a short period of time.

\section{Soliton in a combined harmonic trap and optical lattice\label{Sec:Lattice}}

The classical nonlinear dynamics of a phase kink in a combined harmonic trap and optical lattice
exhibits dynamical instabilities and the soliton may dissipate energy via sound emission even without
quantum and thermal fluctuations \cite{Keverekedis_PRA_2003,Theocharis_MCSim_2005}. The interplay between quantum fluctuations and the dynamical instabilities of the corresponding classical system
was previously studied by us \cite{Martin_unpublished_2009}. We found that the fluctuations can enhance
the effect of instabilities,
in some situations resulting in a surprising {\em reduction} in the position uncertainty of the solitons. Single-shot soliton trajectories displayed splitting and disappearing solitons in an optical lattice, similar to those following collisions, studied in section \ref{Sec:Collisions}.
Here we will focus on solitons that retain their integrity during the evolution. We consider very slow
solitons that would have zero initial velocity in the classical case. Such a system exhibits strong fluctuation-induced effects and we calculate quantum statistical properties of the soliton dynamics,
such as atom number fluctuations in an individual sites, waiting time distributions for the soliton to exit the initial site, atom populations and pair-correlations.

In the simulations of imprinting and dynamics we include in \eref{GPE} the optical lattice potential
$V_{\mbox{\scriptsize exp}}=m\omega^{2}x^{2}/2+sE_{r}\sin^{2}(\pi x/d)$,
where $E_{r}=\hbar^{2}\pi^{2}/2md^{2}$ is the lattice photon recoil energy and $d$ is the lattice period. Here we set $d=\pi l/4$ and $s=1$. After the simulations, in order to translate Wigner distributed stochastic variables to normally ordered expectation values, following the approach in \cite{Isella_PRA_2005,Isella_PRA_2006} we define the ground-state operators $a_{j}$ for the individual lattice sites as
\begin{equation}
a_{j}=\int_{j\mbox{\scriptsize th well}}dx\psi^{*}_{j}(x)\psi_{W}(x,t)
\end{equation}
where $\psi_{j}$ is the Gaussian ground-state wavefunction (Wannier function) of the $j$th well and $\psi_{W}(x,t)$ is the numerically integrated full Wigner wavefunction. The population of the $j$th site is thus $\langle a_{j}^{\dagger}a_{j}\rangle=\langle a_{j}^{*}a_{j}\rangle_{W}-1/2$, and the number fluctuations in the $j$th site
\begin{equation}
\delta n_{j}=\left[\langle(a_{j}^{\dagger}a_{j})^{2}\rangle-\langle a_{j}^{\dagger}a_{j} \rangle^{2} \right]^{1/2}=\left[\langle(a_{j}^{*}a_{j})^{2}\rangle_{W}-\langle a_{j}^{*}a_{j} \rangle_{W}^{2}-\frac{1}{4} \right]^{1/2}.
\end{equation}
The pair-correlation function between the $k$th and the 0th site (the central site in which the soliton is imprinted) reads
\begin{eqnarray}
&g_{2}(0,k)=\langle a_{0}^{\dagger}a_{k}^{\dagger} a_{k}a_{0}\rangle\\ \nonumber
&=\langle a_{0}^{*}a_{k}^{*} a_{k}a_{0}\rangle_{W}-\frac{1}{2}\left(\langle a_{0}^{*}a_{0}\rangle_{W}+\langle a_{k}^{*}a_{k}\rangle_{W} -\frac{1}{2} \right)-\delta_{0k}\left[\langle a_{k}^{*}a_{k}\rangle_{W}-\frac{1}{4}\right].
\end{eqnarray}

\begin{figure}[tbp]
\centering
\includegraphics[width=12.0cm]{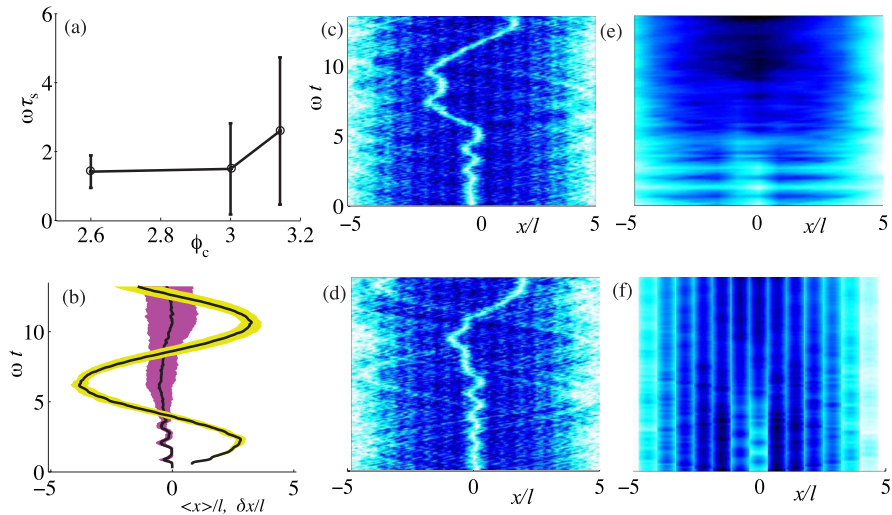}
\caption{Solitons in a combined harmonic trap and optical lattice, with $N_{\mbox{\scriptsize tot}}=900$ and $T=0$. (a) The quantum expectation value of the time $\tau_{s}$ that it takes for a soliton to exit the central lattice site and its quantum uncertainty (error bar) vs the phase jump $\phi_{c}$; (b) quantum expectation values for the soliton position $\langle x(t) \rangle$ (lines) and its uncertainty $\delta x$ (shaded areas) for $\phi_{c}=\pi$ (magenta) and $\phi_{c}=2$ (yellow); (c)-(d) the Wigner density $|\psi_{W}(x,t)|^{2}$ for individual realizations of soliton dynamics with $\phi_{c}=\pi$; (e) the correlation function $g_{2}(0,x)$; (f) the density expectation value in the lowest band. (e) and (f) both show distributions becoming flat due to randomly drifting solitons along the lattice without solitons in individual realizations disappearing.} \label{Pi_Plots}
\end{figure}

We apply an imprinting potential that would, in the classical (GPE) case, prepare a phase kink of $\phi_{c}=\pi$ (a soliton with zero velocity). This state is unstable \cite{Keverekedis_PRA_2003}, as small oscillations in a lattice site become amplified by sound emission so the soliton can escape from the central site and begin to drift along the lattice \cite{Theocharis_MCSim_2005}. Quantum  fluctuations seed this soliton motion as demonstrated in figure \ref{Pi_Plots}(e)-(f) where the solitons exit the central cite at different times and in different directions. Figure \ref{Pi_Plots}(a) shows that the quantum uncertainty in the time $\tau_{s}$ that it takes for a soliton to exit the central site becomes comparable to the quantum expectation value $\tau_{s}$.

The population expectation value for the central lattice site $\langle n_{0} \rangle$ initially shows a soliton oscillating at the frequency of the lattice site, indicating overlapping oscillations between this site and the adjacent sites. This is followed by a rapid increase in central site occupation $\langle n_{0} \rangle$ after time $t\simeq4/\omega$ when there is a high probability that the soliton has left the site. This behaviour is reflected in the expectation value of the atom density in the lowest energy band at $x=0$ [figure \ref{Pi_Plots}(f)]. We also find an initial oscillatory behaviour followed by a rapid increase in the pair correlation function [figure \ref{Pi_Plots}(e)]. Previous studies of solitons in lattices with small atomic occupations \cite{Mishmash_PRL_2009,Mishmash_unpublished_2009} cited such an increase as evidence of filling of a soliton core in individual realizations, but here in large atom number systems we find a similar effect caused by solitons drifting along the lattice, while a soliton core in none of the individual realizations is significantly filled.
\begin{figure}[tbp]
\centering
\includegraphics[width=12.0cm]{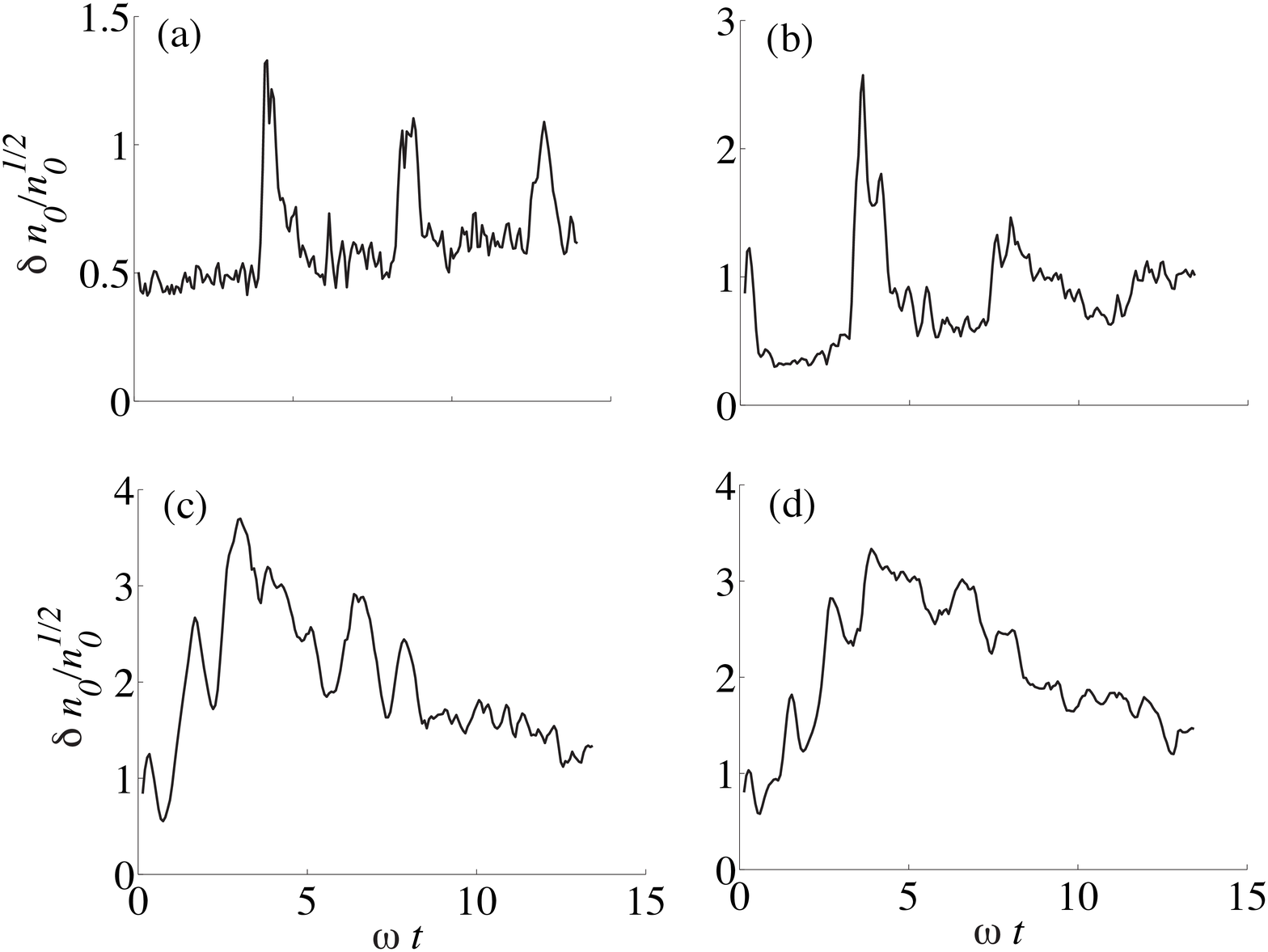}
\caption{Atom number squeezing parameter in the central lattice site for a soliton in a combined harmonic trap and optical lattice, with $N_{\mbox{\scriptsize tot}}=900$ and $T=0$. The phase notch across the soliton is imprinted by a potential which in the corresponding classical system imprints the phases (a) $\phi_{c}=1.3$ (fast soliton), (b) $\phi_{c}=2.6$, (c) $\phi_{c}=3.0$ (d) $\phi_{c}=\pi$ (classical soliton has zero velocity). Soliton crossing $x=0$ is identified by super-Poisson atom number fluctuations. At lower soliton velocities, the sharp peaks are broadened due to the quantum uncertainty in the soliton crossing time.} \label{Num_Fluc_Plots}
\end{figure}

As well as affecting the expectation values $\langle n_{0}\rangle $ and $g_{2}$, the drifting behaviour of solitons has dramatic effects on the atom number fluctuations. Figure \ref{Num_Fluc_Plots} shows the number squeezing parameter in the central lattice site $\delta n_{0}/n_{0}^{1/2}$ corresponding to different $\phi_{c}$ generated by changing the imprinting time $\tau$ in the potential (\ref{Eq:VInt}). When the fluctuations obey a Poisson distribution, $\delta n_{0}/n_{0}^{1/2}=1$. The atom number fluctuations of a ground-state BEC in an optical lattice are sub-Poissonian (squeezed), such
that $\delta n_{0}/n_{0}^{1/2}<1$ \cite{Isella_PRA_2005,Isella_PRA_2006}.
For fast solitons, the soliton position uncertainty is small [figure \ref{Pi_Plots} (b)], soliton cores between different realizations overlap, so there are predictable times when the soliton at high probability is not in the central lattice site. During these times number statistics in the site remain squeezed. The crossing of the soliton at $x=0$ can be recognized as a sharp super-Poissonian peak in the dynamics of the squeezing parameter. For slower solitons, the position uncertainty of the soliton is so large that there are large atom number fluctuations in the central lattice site due to fluctuations in the crossing times of the solitons at $x=0$.

\ack We acknowledge financial support from the EPSRC.

\section*{References}

\bibliography{Bib_TWA}

\begin{thebibliography}{10}

\bibitem{Morsch}
Morsch O and Oberthaler M 2006, {\em Rev. Mod. Phys.} {\bf 78}, 179.

\bibitem{Folman_AMOP_2002}
Folman R, Kr{\"u}ger P, Schmiedmayer J, Denschlag J and Henkel C  2002
  {\em Adv. Atom. Mol. Phys.} {\bf 48},~263.

\bibitem{Tolra_PRL_2004}
 Laburthe Tolra B, O'Hara K M,  2004 {\em Phys. Rev. Lett.}
{\bf 92}, 190401.

\bibitem{Kinoshita_Nature_2006}
Kinoshita T, Wenger T and Weiss D~S  2005 {\em Nature} {\bf 440},~900.

\bibitem{Hofferberth_NPhys_2008}
Hofferberth S, Lesanovsky I, Schumm T, Imambekov A, Gritsev. V, Demler E
  and Schmiedmayer J  2008 {\em Nature Phys.} {\bf 4},~489.

\bibitem{Paredes_Nature_2004}
 Paredes B, Widera A, Murg V, Mandel O, F\"{o}lling S,
Cirac I, Shlyapnikov G V, H\"{a}nsch T W and Bloch I 2004 {\em Nature}
{\bf 429}, 277.

\bibitem{Fertig_PRL_2005}
Fertig C~D, O'Hara K~M, Rolston S~L, Phillips W~D and Porto J~V  2005
  {\em Phys. Rev. Lett.} {\bf 94},~120403.

\bibitem{Mun_PRL_2007}
Mun J, Medley P, Campbell G~K, Marcassa L~G, Pritchard D~E and
  Ketterle W  2007 {\em Phys. Rev. Lett.} {\bf 99},~150604.

\bibitem{Isella_PRA_2005}
Isella L and Ruostekoski J  2005 {\em Phys. Rev. A} {\bf 72},~011601(R).

\bibitem{Isella_PRL_2005}
Ruostekoski J and Isella L  2005 {\em Phys. Rev. Lett.} {\bf
  95},~110403.

\bibitem{Isella_PRA_2006}
Isella L and Ruostekoski J  2006 {\em Phys. Rev. A} {\bf 74},~063625.

\bibitem{POL03c}
Polkovnikov A and Wang D W, 2004 {\em Phys. Rev. Lett.} 
{\bf 93}, 070401.

\bibitem{Drummond_EPhysLett_1993}
Drummond P~D and Hardman A~D  1993 {\em Europhys. Lett.} {\bf
  21},~279.

\bibitem{Steel_PRA_1998}
Steel M~J, Olsen M~K, Plimak L~I, Drummond P~D, Tan S~M, Collett M~J, Walls D~F
  and Graham R  1998 {\em Phys. Rev. A} {\bf 58},~4824.

\bibitem{Sinatra_JPhysB_2002}
Sinatra A, Lobo C and Castin Y  2002 {\em J. Phys. B} {\bf 35},~3599.

\bibitem{Blakie_AdvInPhys_2008}
Blakie P~B, Bradley A~S, Davis M~J, Ballagh R~J and Gardiner C~W  2008
  {\em Adv. in Phys.} {\bf 57},~363.
  
\bibitem{Anglin_NaturePhys_2008}
Anglin J 2008, {\em Nature Phys.} {\bf 4}, 437.

\bibitem{Burger_PRL_1999}
Burger S, Bongs K, Dettmer S, Ertmer W and Sengstock K  1999 {\em
  Phys. Rev. Lett.} {\bf 83},~5198.

\bibitem{Denschlag_Science_2000}
Denschlag J, Simsarian J~E, Feder D~L, Clark C~W, Collins L~A, Cubizolles J,
  Deng L, Hagley E~W, Helmerson K, Reinhardt W~P, Rolston S~L, Schneider B~I
  and Phillips W~D  2000 {\em Science} {\bf 287},~97.

\bibitem{Becker_Nature_2008}
Becker C, Stellmer S, Soltan-Panahi P, Dorscher S, Baumert M, Richter E~M,
  Kronjager J, Bongs K and Sengstock K  2008 {\em Nature Phys.} {\bf
  4},~496.

\bibitem{Martin_unpublished_2009}
Martin A~D and Ruostekoski J  2009 {\em arXiv:0909.2621} .

\bibitem{optsoliton}
 Corney J F and  Drummond P D 2001 {\em J. Opt. Soc. Am. B}
{\bf 18}, 153.

\bibitem{Fedichev_PRA_1999}
Fedichev P~O, Muryshev A~E and Shlyapnikov G~V  1999 {\em Phys. Rev.
  A} {\bf 60},~3320.

\bibitem{Jackson_PRA_2007}
Jackson B, Proukakis N~P and Barenghi C~F  2007 {\em Phys. Rev. A}
  {\bf 75},~051601.

\bibitem{Cockburn_unpublished_2009}
Cockburn S~P, Nistazakis H~E, Horikis T~P, Keverekidis P~G, Proukakis N~P
  and Frantzeskakis D~J  2009 {\em arXiv:0909.1660v2} .

\bibitem{Dziarmaga_PRA_2002}
Dziarmaga J,  Karkuszewski  Z P and Sacha K  2002 {\em Phys. Rev. A} {\bf 66},~043615.

\bibitem{Dziarmaga_PRA_2002_b}
Dziarmaga J  and Sacha K 2002 {\em Phys. Rev. A} {\bf 66},~043620.

\bibitem{Dziarmaga_JPhysB_2003}
Dziarmaga J, Karkusszewski Z~P and Sacha K  2003 {\em J. Phys. B} {\bf
  36},~1217.

\bibitem{Dziarmaga_PRA_2004}
Dziarmaga J  2004 {\em Phys. Rev. A} {\bf 70},~063616.

\bibitem{Damski_PRA_2006}
Damski B  2006 {\em Phys. Rev. A} {\bf 73},~043601.

\bibitem{Mishmash_PRL_2009}
Mishmash R~V and Carr L~D  2009 {\em Phys. Rev. Lett.} {\bf
  103},~140403.
  
\bibitem{Mishmash_unpublished_2009}
Mishmash R~V, Danshita I, Clark C W and Carr L D 2009 {\em Phys. Rev. A} {\bf 80}, 053612.

\bibitem{KRU09}
Krutitsky K V, Larson J and Lewenstein M, 2009 {\em arXiv:0907.0625}

\bibitem{Negretti_PRA_2008_b}
Negretti A, Henkel C and M{\o}lmer K  2008 {\em Phys. Rev. A}
  {\bf 78},~023630.

\bibitem{Dutton_Science_2001}
Dutton Z, Budde M, Slowe C and Hau L~V  2001 {\em Science} {\bf
  293},~663.
  
\bibitem{Ginsberg_PRL_2005}
Ginsberg NS, Brand J and  Hau L V  2005 {\em Phys. Rev. Lett.} {\bf
 94},~040403.

\bibitem{Jo_PRL_2007}
Jo G~B, Choi J~H, Christensen C~A, Pasquini T~A, Lee Y~R, Ketterle W
  and Pritchard D~E  2007 {\em Phys. Rev. Lett.} {\bf 98},~180401.

\bibitem{Engels_PRL_2007}
Engels P and Atherton C  2007 {\em Phys. Rev. Lett.} {\bf 99},~160405.

\bibitem{Kivshar_PhysRep_1998}
Kivshar Y S and Luther-Davies B 1998 {\em Phys. Rep.} {\bf 298},~81.

\bibitem{imprint} Ruostekoski J and Dutton Z 2005, {\em Phys. Rev. A} {\bf 72}, 063626.

\bibitem{Stellmer_PRL_2008}
Stellmer S, Becker C, Soltan-Panahi P, Rishter E~M, Dorsher S, Baumert M,
  Kronjager K, Bongs K and Sengstock K  2008 {\em Phys. Rev. Lett.}
  {\bf 101},~120406.
  
\bibitem{Shomroni_NPhys_2009}
Shomroni I, Lahoud E, Levy S and Steinhauer J  2009 {\em Nature Phys.}
  {\bf 5},~193.

\bibitem{Weller_PRL_2008}
Weller A, Ronzheimer J~P, Gross C, Esteve J, Oberthaler M~K, Frantzeskakis D~J,
  Theocharis G and Kevrekedis P~G  2008 {\em Phys. Rev. Lett.} {\bf
  101},~130401.
  
\bibitem{Busch_PRL_2000}
Busch Th and Anglin J~R  2000 {\em Phys. Rev. Lett.} {\bf 84},~2298.

\bibitem{Anderson_PRL_2001}
Anderson B~P, Haljan P~C, Regal C~A, Feder D~L, Collins L~A, Clark C~W
  and Cornell E~A  2001 {\em Phys. Rev. Lett.} {\bf 86},~2926.

\bibitem{Muryshev_PRA_1999}
Muryshev A E, van Linden van den Heuvell H B and Shlyapnikov G V 
 1999 {\em Phys. Rev. A} {\bf 60},~R2668.

\bibitem{Huang_PRA_2001}
Huang G,  Velarde M G and  Makarov V A 2001  {\em Phys. Rev. A} {\bf 64},~013617.

\bibitem{Zakharov_JETP_1973}
Zakharov V E and Shabat A B 1973   {\em Zh. Eksp. Teor. Fiz.} {\bf 64}, 1627 [{\em Sov. Phys. JETP} {\bf 37}, 823]

\bibitem{Kheruntsyan_PRL_2003}
 Kheruntsyan K V, Gangardt D M, Drummond P D and  Shlyapnikov G V 2003 {\em Phys. Rev. Lett.} {\bf 91}, 040403.

\bibitem{Mora_PRA_2003}
Mora C and Castin Y  2003 {\em Phys. Rev. A} {\bf 67},~053615.

\bibitem{Shrestha_PRA_2009}
Shrestha U, Javanainen J and Ruostekoski J  2009 {\em Phys. Rev. A} {\bf 79},~043617.

\bibitem{Law_PRA_2003}
Law C~K  2003 {\em Phys. Rev. A} {\bf 68},~015602.

\bibitem{Martin_PRL_2007}
Martin A~D, Adams C~S and Gardiner S~A  2007 {\em Phys. Rev. Lett.}
  {\bf 98},~020402.
  
\bibitem{Ruostekoski_PRA_2001}
Ruostekoski J, Kneer B, Schleich W P and Rempe G 2001 {\em Phys. Rev. A} {\bf 63},~043613.

\bibitem{Keverekedis_PRA_2003}
Kevrekidis P G, Carretero-Gonz‡lez R,  Theocharis G, Frantzeskakis D J and Malomed B A 2003, {\em Phys. Rev A} {\bf 68}, 035602.

\bibitem{Theocharis_MCSim_2005}
Theocharis G, Frantzeskakis D J, Kevrekidis P G, Carretero-Gonz‡lez R and Malomed B A 2005 {\em Math. Comput. Simulat.} {\bf 69}, 537.


\end{thebibliography}

\end{document}